\documentclass[12pt]{iopart}
\usepackage{bbold}
\usepackage{iopams}
\usepackage{graphicx}
\usepackage{float}
\newcommand{\diag}{\mathop{\mathrm{diag}}}

\begin{document}

\title[]{Invariant Quantum Discord in Qubit-Qutrit Systems under Local Dephasing}

\author[]{G. Karpat and Z. Gedik}
\address{Faculty of Engineering and Natural Sciences, Sabanci University, Tuzla, Istanbul 34956, Turkey}
\ead{gkarpat@sabanciuniv.edu}

\begin{abstract}
We investigate the dynamics of quantum discord and entanglement for a class of mixed qubit-qutrit states assuming that only the qutrit is under the action of a dephasing channel. We demonstrate that even though the entanglement in the qubit-qutrit state disappears in a finite time interval, partial coherence left in the system enables quantum discord to remain invariant throughout the whole time evolution.
\end{abstract}

\section{Introduction}
Entanglement is not only considered as one of the most profound traits of quantum mechanics but also widely regarded as the central resource of quantum information science [1]. Although most of the quantum computation and communication processes rely on entanglement, it has been revealed that it is not the only kind of resource responsible for the computational speed-up in quantum information tasks. Indeed, no entanglement is needed for the model of deterministic quantum computation with one qubit, where more general non-classical correlations is accountable for the efficiency of computation [2-4]. These non-classical correlations, that cannot be captured by entanglement measures, is characterized by quantum discord [5,6]. In recent years, quantum discord has been used for analyzing a wide range of problems related to open quantum system dynamics, quantum phase transitions and quantum algorithms [7].

When quantum systems interact with their surroundings, most fundamental quantum features such as the existence of quantum superpositions are irreversibly destroyed through a process known as decoherence [8]. A remarkable consequence of this system-environment interaction is the total disappearance of entanglement between the parts of a composite system in finite time, a phenomenon known as entanglement sudden death [9]. Werlang et al. have comparatively investigated the dynamics of entanglement and quantum discord in Markovian environments and shown that, unlike entanglement, quantum discord is immune to sudden death [10]. On the other hand, Mazzola et al. have discovered the existence of a sudden transition between classical and quantum loss of correlations for a special class of Bell diagonal states under local dephasing noise [11]. This phenomenon implies that there exists a finite time interval, in which only classical correlations decay and quantum discord is frozen despite the presence of a noisy environment. Moreover, Haikka et al. have demonstrated that quantum discord might get forever frozen at a positive value depending on the initial state when both qubits locally interact with non-Markovian purely dephasing environments [12].

A comprehensive analysis of the dynamics of various quantum and classical correlation measures for qubit-qutrit systems interacting with multilocal and global dephasing environments has been carried out in Ref. [13]. It has been shown that the phenomenon of sudden transition between classical and quantum decoherence can also be observed in certain families of qubit-qutrit states under multilocal dephasing noise. In this work, we explore the dynamics of quantum discord and entanglement for qubit-qutrit states assuming that only the qutrit is under the action of a Markovian dephasing channel and the qubit is not affected by noisy environment. We show that although the entanglement between the qubit and the qutrit, as quantified by negativity, evaporates in a finite time interval, quantum discord does not feel the noisy environment and remains invariant during the whole dynamics.

\section{Measures of Quantum Correlations}

In this section, we introduce the measures of quantum correlations used in our study. Despite the fact that the quantification of entanglement is completely understood for the quantum systems composed of two-qubits [14,15], the problem is still widely open for the case of higher dimensional mixed states. Negativity is a reliable measure of entanglement that can be computed straightforwardly for an arbitrary bipartite state regardless of its dimension, provided that the state has a negative partial transpose. It is in general not possible to conclude whether a positive partial transpose state is separable or not. However, it has been proved that all positive partial transpose states of qubit-qubit and qubit-qutrit systems are separable [16,17]. Hence, negativity completely characterizes the qubit-qutrit entanglement. For a given bipartite system $\rho^{AB}$, negativity is calculated as twice the absolute sum of the negative eigenvalues of partial transpose of $\rho^{AB}$ with respect to the smaller dimensional system,
\begin{equation}
N(\rho^{AB})=\sum_{i}|\eta_{i}|-\eta_{i},
\end{equation}
where $\eta_{i}$ are all of the eigenvalues of the partially transposed density matrix $(\rho^{AB})^{T_{A}}$.

The quantum mutual information quantifies the total amount of classical and quantum correlations contained in a quantum state, and can be evaluated as
\begin{equation}
I(\rho^{AB})=S(\rho^{A})+S(\rho^{B})-S(\rho^{AB}),
\end{equation}
where $S(\rho)=-Tr(\rho \textmd{log}_{2} \rho)$ is the von-Neumann entropy with $\rho^{AB}$ and $\rho^{k}$ $(k=A, B)$ being the density matrix of the total system and reduced density matrix of subsystems, respectively. The amount of classical correlations present in a quantum state can be measured by [5,6]
\begin{equation}
C(\rho^{AB})= S(\rho^{B})-\min_{\{\Pi_{k}^{A}\}}\sum_{k}p_{k}S(\rho_{k}^{B}),
\end{equation}
where $\{\Pi_{k}^{A}\}$ defines a set of orthonormal projection operators, acting on the subsystem $A$ and $\rho_{k}^{B}=Tr_{A}((\Pi_{k}^{A} \otimes I^{B})\rho^{AB})/p_{k}$ is the remaining state of the subsystem $B$ after obtaining the outcome $k$ with the probability $p_{k}=Tr((\Pi_{k}^{A} \otimes I^{B})\rho^{AB})$. We will evaluate $C(\rho^{AB})$ for qubit-qutrit states under the assumption that the measurement is performed on the qubit part of the composite system. The measurement operators $\{\Pi_{1}^{A}, \Pi_{2}^{A}\}$ can be parameterized as
\begin{eqnarray}
\Pi_{1}^{A}=\frac{1}{2}\left(I^{A}_{2}+\sum_{j=1}^{3}n_{j}\sigma_{j}^{A}\right),  \nonumber \\
\Pi_{2}^{A}=\frac{1}{2}\left(I^{A}_{2}-\sum_{j=1}^{3}n_{j}\sigma_{j}^{A}\right),
\end{eqnarray}
where $\sigma_{j}(j=1,2,3)$ are the Pauli spin matrices and $n=(\sin\theta\cos\phi,\sin\theta\sin\phi,\cos\theta)^{T}$ is a unit vector on the Bloch sphere with $\theta \in [0,\pi)$ and $\phi \in [0,2\pi)$. Then, quantum discord [5], that measures the amount of non-classical correlations, is  defined as the difference between total and classical correlations,
\begin{equation}
D(\rho^{AB})=I(\rho^{AB})-C(\rho^{AB}).
\end{equation}
Considering the fact that the optimization process involved in the calculation of quantum discord might get complicated depending on the quantum state, there exists no closed analytical expression even for most general case of two-qubit states. For the simple qubit-qutrit states that we consider in our work, we will calculate the quantum discord via numerical minimization over two independent real parameters $\theta$ and $\phi$.

\section{Time Invariant Quantum Discord}

Dynamical evolution of an open quantum system can be described in terms of a completely positive trace preserving linear map $\mathcal{E}$ acting on the state space of the considered system. It has been shown that, for every completely positive trace preserving map, there exists a corresponding operator-sum representation also known as the Kraus representation [18,19]. For an arbitrary initial density matrix, the effect of the linear quantum map $\mathcal{E}$ is represented by the collective action of a set of (non-unique and not necessarily unitary) Kraus operators $\{K_{i}\}$ as
\begin{equation}
\rho(t) = \mathcal{E}(\rho(0)) = \sum_{i=1}^{N}K_{i}(t)\rho(0)K_{i}^{\dagger}(t),
\end{equation}
where the Kraus operators $K_{i}$ satisfy the normalization condition
\begin{equation}
\sum_{i=1}^{N}K_{i}^{\dagger}(t)K_{i}(t)=I,
\end{equation}
for all values of $t$. The operator-sum approach is quite general since the set of Kraus operators $\{K_{i}\}$ intrinsically contains the entire information about the environment without explicitly considering its detailed properties.

We now evaluate the dynamics for hybrid qubit-qutrit states under the assumption that only the qutrit is interacting with Markovian dephasing environment and the qubit is protected. The operator-sum representation of the considered qutrit dephasing channel can be described by a set of Kraus operators [20],
\begin{eqnarray}
\fl
M_{1}= \diag(1,\gamma(t),\gamma(t)),
\qquad
M_{2}= \diag(0,\omega(t),0),
\qquad
M_{3}= \diag(0,0,\omega(t)),
\end{eqnarray}
where $\gamma(t)=e^{-\Gamma t/2}$ and $\omega(t)=\sqrt{1-\gamma^{2}(t)}$ with $\Gamma$ denoting the decay rate. This specific channel is chosen so that the rate of dephasing between the ground state and each of the two excited states are the same. We also note that, as a general property, dephasing channels do not change the quantum state populations and thus only produces a loss of coherence without altering the energy of the system. Having defined the decoherence channel for a single qutrit, we can obtain the time evolution of an arbitrary initial qubit-qutrit system $\rho(0)$ under local dephasing of the qutrit as
\begin{equation}
\rho(t) =\sum_{i=1}^{3}(I_{2} \otimes M_{i})\rho(0)(I_{2} \otimes M_{i})^{\dagger},
\end{equation}
where $I_{2}$ denotes the $2\times2$ identity matrix acting on the qubit part of the composite system. The resulting time-evolved density matrix in the product basis $\lbrace \vert ij\rangle : i=0, 1, j= 0, 1, 2 \rbrace$ can be then written as
\begin{equation}
\rho(t)
= \left(\begin{array}{cccccc}
\rho_{11} & \rho_{12} \gamma & \rho_{13} \gamma & \rho_{14} & \rho_{15} \gamma & \rho_{16} \gamma\\
\rho_{21} \gamma & \rho_{22} & \rho_{23} \gamma^{2} & \rho_{24} \gamma & \rho_{25} & \rho_{26} \gamma^{2}\\
\rho_{31} \gamma & \rho_{32} \gamma^{2} & \rho_{33} & \rho_{34} \gamma & \rho_{35} \gamma^{2} & \rho_{36} \\
\rho_{41} & \rho_{42} \gamma& \rho_{43} \gamma & \rho_{44} & \rho_{45} \gamma & \rho_{46} \gamma\\
\rho_{51} \gamma & \rho_{52}  & \rho_{53} \gamma^{2}& \rho_{54} \gamma & \rho_{55} & \rho_{56} \gamma^{2}\\
\rho_{61} \gamma & \rho_{62} \gamma^{2} & \rho_{63} & \rho_{64} \gamma & \rho_{65} \gamma^{2} & \rho_{66}\\
\end{array}\right).
\end{equation}
In the following, we analyze the time evolution of quantum correlations for a one-parameter family of entangled qubit-qutrit mixed states
\begin{eqnarray}
\fl
\rho = \frac{p}{2}(|00\rangle\langle00|+|01\rangle\langle01|+|12\rangle\langle12|+
|11\rangle\langle11|+|01\rangle\langle11|+|11\rangle\langle01|+|00\rangle\langle12| \nonumber \\
+|12\rangle\langle00|)+\frac{1-2p}{2}(|02\rangle\langle02|+
|02\rangle\langle10|+|10\rangle\langle02|+|10\rangle\langle10|),
\end{eqnarray}
where $p\in[0,0.5]$ and $\rho$ is separable only for $p=1/3$.

In Fig. 1(a), we present our results on the dynamics of negativity and quantum discord as a function of the dimensionless parameter $\Gamma t$ for $p=0.15$. We notice that although the coherence in the qubit-qutrit system is only partially lost, entanglement as quantified by negativity suffers a sudden death and disappears after a certain finite time. On the other hand, quantum discord remains frozen for a while but then when a critical instance is reached, it decays to a finite non-zero value. The survival of quantum discord at the asymptotic limit ($t\rightarrow\infty$) is not unexpected since the quantum state is still partially coherent and almost all quantum states have non-classical correlations [21]. Regardless, Fig. 2(a) displays a curious behavior of the correlations for $p=0.23$. In this case, we observe that even if the negativity evaporates quickly due to sudden
\begin{figure}[ht!]
\begin{center}
\includegraphics[scale=0.67]{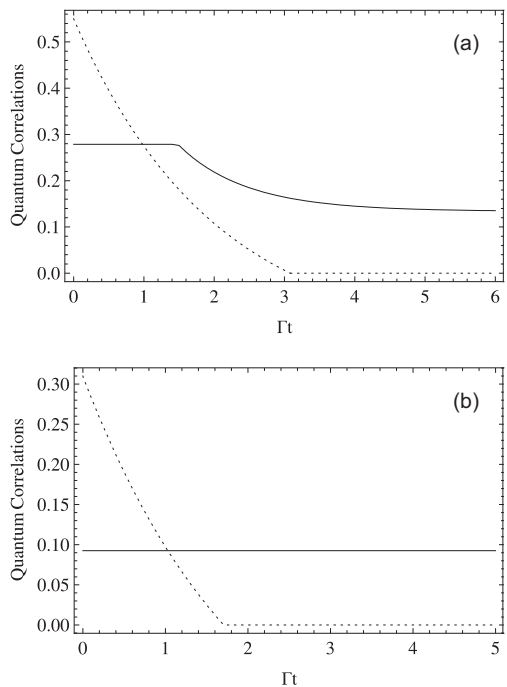}
\caption{Dynamics of negativity (dotted line) and quantum discord (solid line) as a function of the dimensionless parameter $\Gamma t$ for $p=0.15$ (a) and $p=0.23$ (b).}
\end{center}
\end{figure}
death, the partial coherence left in the qubit-qutrit system enables quantum discord to remain invariant during the whole time evolution. It is important to emphasize that this is a rather surprising feature of non-classical correlations that are more general than entanglement.

\section{Conclusion}

In summary, we have studied time evolutions of entanglement and quantum discord for a certain class of hybrid qubit-qutrit states assuming that the qutrit is interacting with a Markovian dephasing environment. We have shown that the entanglement between the qubit and the qutrit vanishes exhibiting a sudden death, despite the partial coherence left in the system. Nonetheless, for some special initial conditions, quantum discord may not feel the effect of the noisy environment and remain unchanged at all times. Since there is no available analytical expression for the quantum discord of qubit-qutrit states, we are not able to deduce the general form of the quantum states having this property. Finally, whether this kind of dynamics can be also observed for two-qubit states or it can only be seen in higher dimensions, remains as an open question.

\section*{Acknowledgements}
This work has been partially supported by the Scientific and Technological Research Council of Turkey (TUBITAK) under Grant 111T232.


\section*{References}
\begin{thebibliography}{100}
\bibitem{1} Horodecki R, Horodecki P, Horodecki M and Horodecki K 2009 \textit{Rev. Mod. Phys.} \textbf{81} 865
\bibitem{2} Knill E and Laflamme R 1998 \textit{Phys. Rev. Lett.} \textbf{81} 5672
\bibitem{3} Datta A, Shaji A and Caves C M 2008 \textit{Phys. Rev. Lett.} \textbf{100} 050502
\bibitem{4} Lanyon B P, Barbieri M, Almeida M P and White A G 2008 \textit{Phys. Rev. Lett.} \textbf{101} 200501
\bibitem{5} Ollivier H and Zurek W H 2001 \textit{Phys. Rev. Lett.} \textbf{88} 017901
\bibitem{6} Henderson L and Vedral V 2001 \textit{J. Phys. A} \textbf{34} 6899
\bibitem{7} Modi K, Brodutch A, Cable H, Paterek T and Vlatko V \textit{Preprint} arXiv:1112.6238v2
\bibitem{8} Zurek W H 2003 \textit{Rev. Mod. Phys.} \textbf{75} 715
\bibitem{9} Yu T and Eberly J H 2004 \textit{Phys. Rev. Lett.} \textbf{93} 140404
\bibitem{10} Werlang T, Souza S, Fanchini F F and Boas C J V 2009, \textit{Phys. Rev. A} \textbf{80} 024103
\bibitem{11} Mazzola L, Piilo J and Maniscalco S 2010 \textit{Phys. Rev. Lett.} \textbf{104} 200401
\bibitem{12} Haikka P, Johnson T H and Maniscalco S \textit{Preprint} arXiv:1203.6469v2
\bibitem{13} Karpat G and Gedik Z 2011 Phys. Lett. A \textbf{375} 4166
\bibitem{14} Wootters W K 1998 \textit{Phys. Rev. Lett.} \textbf{80} 2245
\bibitem{15} Vidal G and Werner R F 2002 \textit{Phys. Rev. A} \textbf{65} 032314
\bibitem{16} Peres A 1996 \textit{Phys. Rev. Lett.} \textbf{77} 1413
\bibitem{17} Horodecki M, Horodecki P and Horodecki R 1996 \textit{Phys. Lett. A} \textbf{223} 1
\bibitem{18} Kraus K 1983 \textit{States, Effects and Operations: Fundamental Notions of Quantum Theory} (Verlag, Berlin: Springer)
\bibitem{19} Choi M D 1975 \textit{Lin. Alg. and Appl.} \textbf{10} 285
\bibitem{20} Ann K and Jaeger G 2008 \textit{Phys. Lett. A} \textbf{372} 579
\bibitem{21} Ferraro A, Aolita L, Cavalcanti D, Cucchietti F M and Acin A 2010 \textit{Phys. Rev. A} \textbf{81} 052318


\end{thebibliography}
\end{document}